\documentclass[10pt,twocolumn,letterpaper]{article}
\usepackage{cvpr}      
\definecolor{cvprblue}{rgb}{0.21,0.49,0.74}
\usepackage[pagebackref,breaklinks,colorlinks,linkcolor=red,citecolor=cvprblue]{hyperref}
\def\paperID{*****} 
\def\confName{CVPR}
\def\confYear{2025}

\title{DynamicAvatars: Accurate Dynamic Facial Avatars Reconstruction and Precise Editing with Diffusion Models}

\author{
\textbf{Yangyang Qian$^1$} \quad 
\textbf{Yuan Sun$^{\text{\textbf{\textcolor{red}{*}}}2,3,4}$} \quad 
\textbf{Yu Guo$^{\text{\textbf{\textcolor{red}{*}}}2,3,4}$} \quad
\vspace{0.2cm} \\
$ ^1$School of Software Engineering, Xi'an Jiaotong University \\
$ ^2$National Key Laboratory of Human-Machine Hybrid Augmented Intelligence, Xi'an Jiaotong University \\
$ ^3$National Engineering Research Center for Visual Information and Applications, Xi'an Jiaotong University  \\
$ ^4$Institute of Artificial Intelligence and Robotics, Xi'an Jiaotong University
{\fontsize{10}{10}\selectfont}
}

\begin{document}
\twocolumn[{%
\renewcommand\twocolumn[1][]{#1}%
\maketitle
\begin{center}
    \centering
    \captionsetup{type=figure}
    \includegraphics[width=\textwidth]{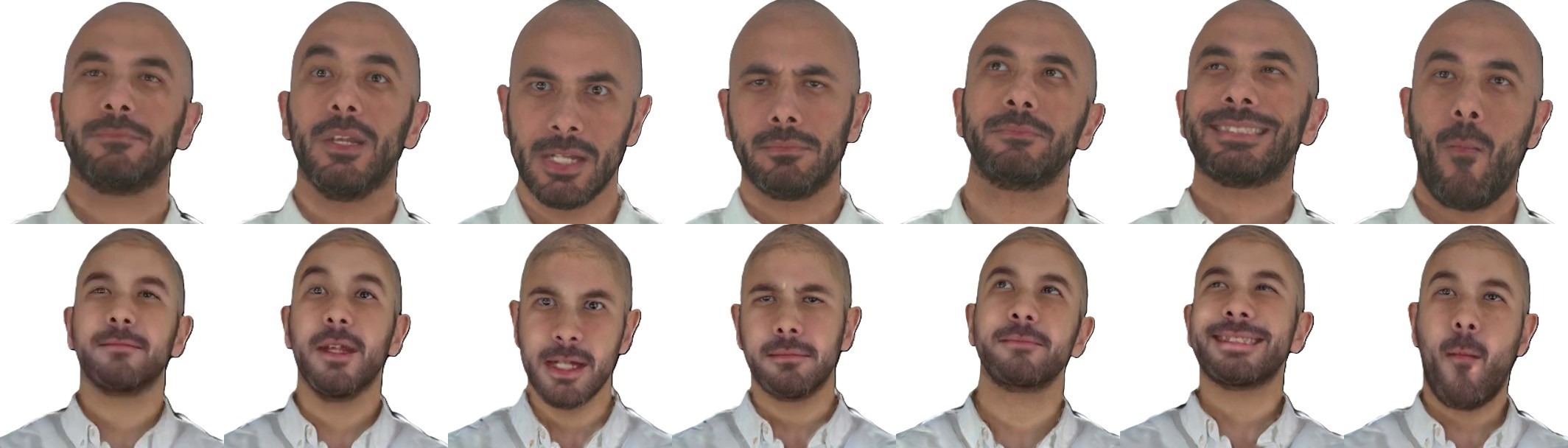}
    \captionof{figure}{Demonstration results of our method. DynamicAvatars is a powerful method which has the ability to render photorealistic images from dynamic models with flexible editing skills. With dual tracking method and LLM guiding prompts, we can easily experience the changing ability include avatar's expression, appearance and accessories. }
    \label{demo}
\end{center}%
}]

\begin{abstract}
Generating and editing dynamic 3D head avatars are crucial tasks in virtual reality and film production. However, existing methods often suffer from facial distortions, inaccurate head movements, and limited fine-grained editing capabilities. To address these challenges, we present DynamicAvatars, a dynamic model that generates photorealistic, moving 3D head avatars from video clips and parameters associated with facial positions and expressions. Our approach enables precise editing through a novel prompt-based editing model, which integrates user-provided prompts with guiding parameters derived from large language models\cite{minaee2024largelanguagemodelssurvey} (LLMs). To achieve this, we propose a dual-tracking framework based on Gaussian Splatting and introduce a prompt preprocessing module to enhance editing stability. By incorporating a specialized GAN algorithm and connecting it to our control module, which generates precise guiding parameters from LLMs, we successfully address the limitations of existing methods. Additionally, we develop a dynamic editing strategy that selectively utilizes specific training datasets to improve the efficiency and adaptability of the model for dynamic editing tasks.
\end{abstract}

\begin{figure*}[htp]
    \centering
    \includegraphics[width=0.7\textwidth]{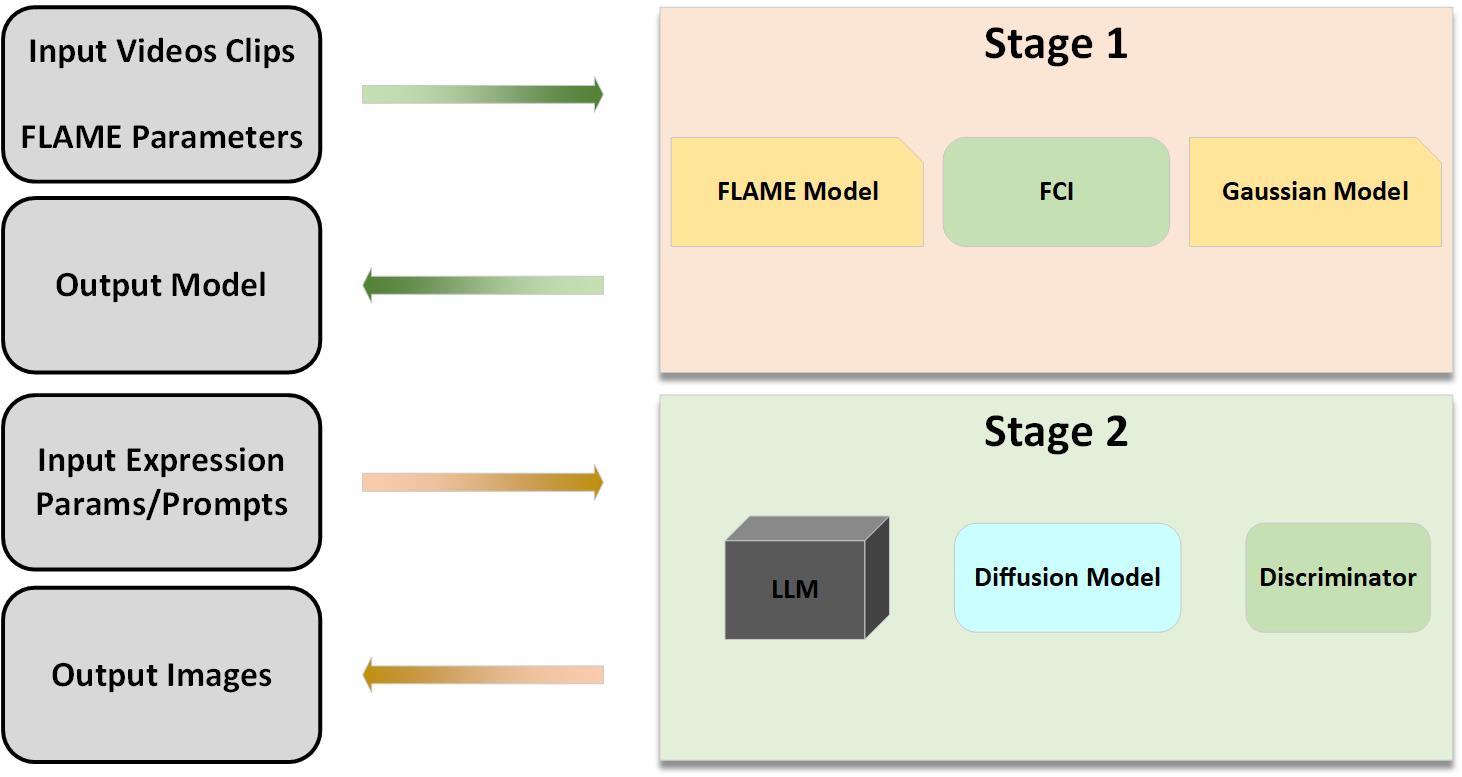}
    \caption{The pipeline of our method. Our pipeline is divided into two main stages: the modeling stage and the editing stage. In modeling stage, video clips and FLAME model parameters are used as input. The objective is to train a model consisting of Gaussian splats, regulated by the FLAME model. The output of this stage is a dynamic Gaussian model capable of accurately representing the head avatar. In the editing stage, expression parameters and guiding prompts are provided as input. A Large Language Model (LLM) is employed to refine and enhance the structure of the prompts, while a discriminator improves the quality of the style-edited images generated by the model. The output of this stage consists of rendered images that reflect the applied edits.}
    \label{fig:pipeline}
\end{figure*}    
\section{Introduction}
\label{sec:intro}

Creating and editing digital avatars of human heads has become a prominent research direction in computer vision, driven by its importance in gaming, film production, and environmental simulation. Efficiently producing and flexibly editing detailed human models is critical for these applications.

Traditional methods utilizing explicit 3D representations, such as point clouds and meshes, often struggle to accurately capture fine geometric details. The inherent complexity of human faces, with their intricate textures, unpredictable poses, and dynamic expressions, further complicates the task of identifying and modeling underlying patterns.

Implicit 3D representation methods have addressed many of these challenges, offering the capability to reconstruct photorealistic human avatars. Neural Radiance Fields \cite{10.1145/3503250} (NeRF) have demonstrated remarkable success by introducing novel rendering pipelines and neural-network-based color storage techniques. NeRF-based approaches like \cite{martinbrualla2020nerfw,pumarola2020d,hedman2021snerg,yu2023dylin,cao2024lightning} have enabled the generation of highly complex scenes and significantly reduced temporal and spatial computation costs. They have also achieved impressive results in reconstructing dynamic scenes and rendering novel views. However, these methods often lack the ability to edit or control facial reconstructions due to limitations in encoding techniques and insufficient use of prior information.

To advance this field, researchers are increasingly turning to 3D Gaussian Splatting \cite{kerbl3Dgaussians} (3DGS) as a promising solution. This method excels in real-time performance and offers a structurally simple framework conducive to editing. Prior works, such as 3DGS-DET \cite{3dgsdet,chen2024omnire} and other works \cite{Yu2023MipSplatting}, have demonstrated the feasibility of 3DGS in 3D object detection and urban scene reconstruction, while recent efforts like Rig3dGS \cite{rivero2024rig3dgs}, Gaussian Editor \cite{chen2023gaussianeditor} and etc.\cite{zhao2024psavatarpointbasedshapemodel,luo2024gaussianhair,shao2024splattingavatar,jiang2024robustdualgaussiansplatting} focus on digital avatars have addressed some challenges by binding meshes to 3D Gaussian splats and optimizing Gaussian cloning and splitting techniques.

Despite these advancements, current models still face significant challenges in achieving precise and flexible editing. One major limitation lies in editing fine facial features and adornments. Existing text-guided image editing models often exhibit low comprehension when interpreting instructions with precise positional details or highly detailed descriptions. Furthermore, editing in dynamic scenes while maintaining real-time performance remains a persistent obstacle.

Our model can reconstruct dynamic digital human head avatars and supports text-based editing of the reconstructed models, as demonstrated in Figure.\ref{demo}. We achieve successful reconstruction by constraining the relative positions of Gaussian splats and meshes while applying a semantic mask to the splats. During the editing phase, we identify all relevant Gaussian splats contributing to the target editing area using a specifically designed strategy. These splats are then refined using an LLM-based editing process to achieve precise modifications. Our pipeline is illustrated in Figure.\ref{fig:pipeline}. As shown, our approach is divided into two stages, enabling the modeling and editing of dynamic 3D scenes using Gaussian Splatting.
Our contributions are mainly reflected in the following points:

1. We propose a dual-tracking Gaussian method to construct a dynamic Gaussian model from a collection of video clips.

2. We design an LLM-based editing process that leverages a GAN framework to enable precise editing of dynamic scenes based on detailed prompts.

\section{Related Work}
\label{sec:formatting}

\subsection{3D Dynamic Head Avatars Editing}

Several techniques have been proposed for tasks involving 3D dynamic scenes. Neural Radiance Fields (NeRF) have introduced a groundbreaking perspective in 3D-aware generation through exceptional volumetric rendering. Efficiency improvements have been achieved by optimizing data organization structures and numerical methods. For example, works such as InstantNGP \cite{mueller2022instant} and Mip-NeRF\cite{barron2021mipnerf} have made significant advancements. However, the implicit information storage approach used by NeRF can be computationally intensive and less efficient compared to explicit methods like voxel grids, point clouds, and Gaussian points.

In our work, we adopt 3D Gaussian Splatting \cite{kerbl3Dgaussians} (3DGS) to balance reconstruction quality in 3D head avatar tasks with effective editing capabilities.

A widely adopted approach for modeling dynamic scenes involves encoding color and volume density information into a Multi-Layer Perceptron (MLP) using spatial-temporal coordinates $(x,y,z,t)$ or relative coordinates in a 4D canonical space like \cite{Wu_2024_CVPR,gafni2020dynamicneuralradiancefields}. More recent methods like \cite{zhou2024headstudio,liu24-GVA}, which attach mesh triangles to 3D Gaussians for explicit editing and fine-tuning, have achieved remarkable results in facial scene reconstruction. Building upon these advancements, we introduce a Gaussian tracing method integrated with mesh adjustment to enable flexible and efficient dynamic editing.

\subsection{LLM-controlled Image Editing Guidance }

Using diffusion models in generating photorealistic images\cite{huang2024diffusionmodelbasedimageediting} has become a standard approach for image generation like \cite{kawar2023imagic,yang2022paint,Yang_2023_CVPR}. Foundational works using prompts to guide the diffusion model, such as InstructPix2Pix \cite{brooks2022instructpix2pix}, achieved this by constructing conditional diffusion models and refining prompts with the help of Large Language Models (LLMs) like ChatGPT. Subsequent studies, such as Prompt2Prompt \cite{hertz2022prompt}, combined structural-aware priors with original text information to enhance control over the generation process.

Recent advancements, such as LLM-grounded Diffusion \cite{lian2023llmgrounded}, have highlighted the pivotal role of LLMs in defining prompts and indirectly influencing the quality of generated results. For instance, a self-correcting, LLM-controlled model proposed in SLD \cite{wu2023self} improved text-to-image alignment by iteratively refining prompts during the generation process.

Building on this line of research, we aim to explore and optimize the editing process further, ultimately contributing to improved precision and accuracy in fine-grained image editing tasks.

\subsection{GAN-based Performance Enhancement}

Generative Adversarial Networks \cite{goodfellow2014generative} (GANs) are widely applied in 3D reconstruction tasks, often in combination with techniques such as Convolutional Neural Networks \cite{oshea2015introductionconvolutionalneuralnetworks} (CNNs) or implicit functions to enhance generated results. These approaches enable the inference of realistic 3D models from limited visual information, enforce geometric coherence across multiple views, and reduce surface texture and geometry blurring.

In the context of digital avatars, GANs have been employed for generating 3D faces and bodies for some time. StyleGAN \cite{karras2019style} is a prominent example, showcasing how GAN-based architectures facilitate expressive face synthesis and demonstrate the potential to generate novel expressions and poses from an initial avatar, while \cite{wang2023diffusiongantraininggansdiffusion} combined GAN with diffusion model which showed positive results. Related works such as EG3D \cite{Chan2021} and GenN2N \cite{liu2024genn2n} provide practical methods for combining GANs with NeRF to improve synthetic novel-view rendering of facial avatars.

In our research, we observed that the original outputs often suffered from inaccurate colors and texture distortions, particularly around sensitive regions such as the eyes and teeth. To address these issues, we implemented a specialized GAN-based generator module designed to specifically target and correct problematic areas, significantly improving the overall quality and realism of the results.
\section{Preliminary}

\subsection{3D Gaussian Splatting}
Gaussian Splatting is based on creating a collection of Gaussian splats, where each splat is characterized by position x and covariance matrix ${\Sigma }$. While the mean of a Gaussian splat is the center point ${\mu }$, for each position $x\in {{\mathbb{R}}^{3}}$ we have:

\begin{equation}\label{eqn-1} 
  G(x)={{e}^{-\frac{1}{2}{{(x-\mu\ )}^{T}}{{\Sigma\ }^{-1}}(x-\mu\ )}}
\end{equation}

Worth to notice that the covariance matrix should be differentiable during optimization process. A matrix decomposition method was introduced to solve this problem as follows:

\begin{equation}\label{eqn-2} 
  \Sigma\ =RS{{S}^{T}}{{R}^{T}}
\end{equation}

where R refers to rotation matrix and S refers to scaling matrix.

Each Gaussian Splat can also be described by a set of physical parameters like position vector ${\mu \in {{\mathbb{R}}^{3}}}$ , scaling vector $s\in {{\mathbb{R}}^{3}}$, rotation quaternion $q\in {{\mathbb{R}}^{4}}$ and color $c\in {{\mathbb{R}}^{3}}$.
The rendering process involves tracing every 3D points along a ray to determine its contribution to the final color of a specific pixel. This is achieved by accumulating the influence of Gaussian splats intersected by the ray, resulting in the rendering formula:

\begin{equation}\label{eqn-3} 
  C=\sum\limits_{i=1}^{N}{{{c}_{i}}{{\alpha }_{i}}\prod\limits_{j=1}^{i-1}{(1-{{\alpha }_{j}})}}
\end{equation}

where ${{c}_{i}}$ represents the color of each 3D point on the trace and ${{\alpha }_{i}}$ is the corresponding density. Note that in an optimized version, we can sort the splats by depth in order to cut off those invisible points. 

\subsection{Precise Edit by LLM}	

\begin{figure*}[htp]
    \centering
    \includegraphics[width=\textwidth]{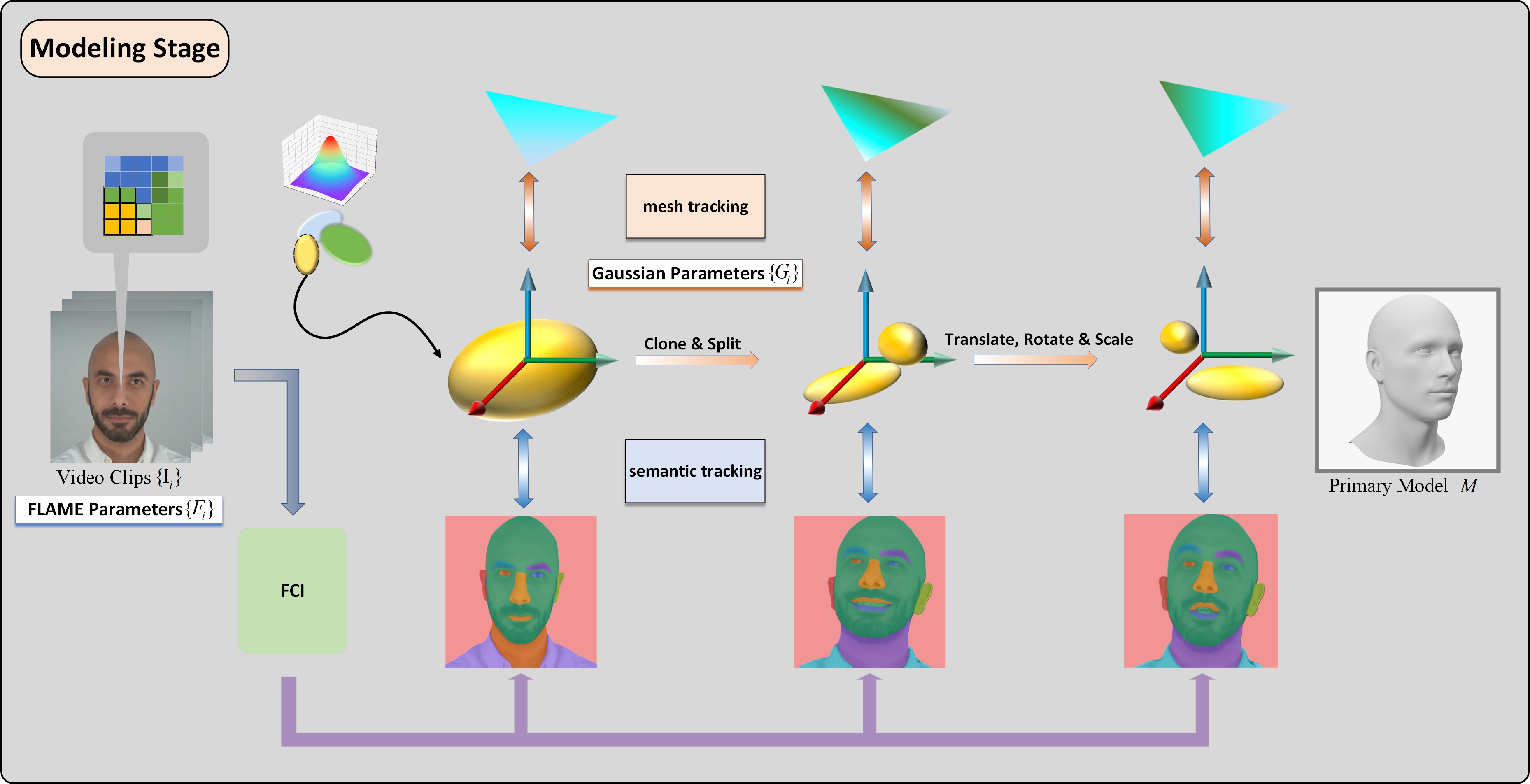}
    \caption{\textbf{Pipeline of the Modeling Stage} In this stage, we utilize a dual tracking method to maintain the relative positions of Gaussian splats, facilitating the editing process in the subsequent stage. For a given set of video clips ${\{I_i\}}$ and the corresponding FLAME parameter set ${\{F_i}\}$, we employ a Facial Component Identifier (FCI) to identify components within the face images. Semantic masks are then used to label the Gaussian splats contributing to each corresponding facial area. Additionally, we bind the Gaussian splats to the mesh of the FLAME model to preserve the spatial structure of the human face. In our experiments, we ultilize }
    \label{fig:stage_1}
\end{figure*}

One of the most important parts of editing stage is to ensure the diffusion model could generate an image which is fully correspond to the text-based instructions. To achieve this, many previous works designed an architecture similar to the follows:

1)\textbf{Define} Let ${ I_{\text{original}}\ \in \mathbb{R}^{H \times W \times C} }$ represent the input image, where ${ H, W, }$ and ${C}$ denote the image height, width, and number of channels, respectively. The user provides a natural language prompt ${T}$, which is a sequence of tokens:  

\begin{equation}\label{eqn-4} 
  T = \{t_1, t_2, \dots, t_n\}, \quad t_i \in \mathcal{V}
\end{equation}

where  ${\mathcal{V}}$ is the vocabulary space, and ${n}$ is the number of tokens in the prompt.

The goal is to generate an edited image  ${I_{\text{edit}}}$  such that the semantics of ${T}$ are incorporated into ${I_{\text{original}}}$ , satisfying:  

\begin{equation}\label{eqn-5} 
  \mathcal{L}_{\text{edit}}(I_{\text{edit}}, T) \to \min
\end{equation}

where  ${\mathcal{L}_{\text{edit}}}$  is a loss function capturing the alignment between  ${I_{\text{edit}}}$  and ${T}$.

2)\textbf{Encode} The input image  ${I_{\text{original}}}$  is passed through an image encoder ${ D_{\text{img}}: \mathbb{R}^{H \times W \times C} \to \mathbb{R}^d}$, producing a latent representation:  

\begin{equation}\label{eqn-6} 
  z_{\text{img}} = D_{\text{img}}(I_{\text{original}})
\end{equation}

where ${z_{\text{img}} \in \mathbb{R}^d}$ is a feature vector in the latent space of dimension ${d}$.

The textual prompt ${T}$ is mapped to the same latent space using a text encoder ${D_{\text{txt}}: \mathcal{V}^n \to \mathbb{R}^d}$:

\begin{equation}\label{eqn-7} 
  z_{\text{txt}} = D_{\text{txt}}(T)
\end{equation}

3)\textbf{Align} To align image and text features, a joint latent space  ${\mathcal{Z}}$ is defined, where: 

\begin{equation}\label{eqn-8} 
  \hat{z}_{\text{img}} = g_{\text{align}}(z_{\text{img}}, z_{\text{txt}})
\end{equation}

Here, ${g_{\text{align}}}$  is a cross-attention mechanism or a shared embedding model that fuses ${ z_{\text{img}}}$ and  ${z_{\text{txt}}}$. The resulting latent representation ${\hat{z}_{\text{img}} \in \mathbb{R}^d}$ captures prompt-specific modifications.

4)\textbf{Edit} The latent code ${\hat{z}_{\text{img}}}$ is updated iteratively to incorporate desired edits. For diffusion-based models, this is achieved via a reverse diffusion process:  

\begin{equation}\label{eqn-9} 
  z_t = z_{\text{img}} + \epsilon_t \cdot \sigma_t, \quad t = 0, 1, \dots, T
\end{equation}

where ${\sigma_t}$  represents noise intensity at step ${t}$, and ${\epsilon_t}$ is sampled from a Gaussian distribution ${\mathcal{N}(0, I)}$.

5)\textbf{Decode} The output image is further refined through optional denoising or super-resolution processes ${\mathcal{S}}$: 

\begin{equation}\label{eqn-10} 
  I_{\text{final}} = \mathcal{S}(I_{\text{edit}})
\end{equation}

\section{Method}

\subsection{Semantic-based Mesh-Gaussian Tracking}

\begin{figure*}[htbp]
    \centering
    \includegraphics[width=\textwidth]{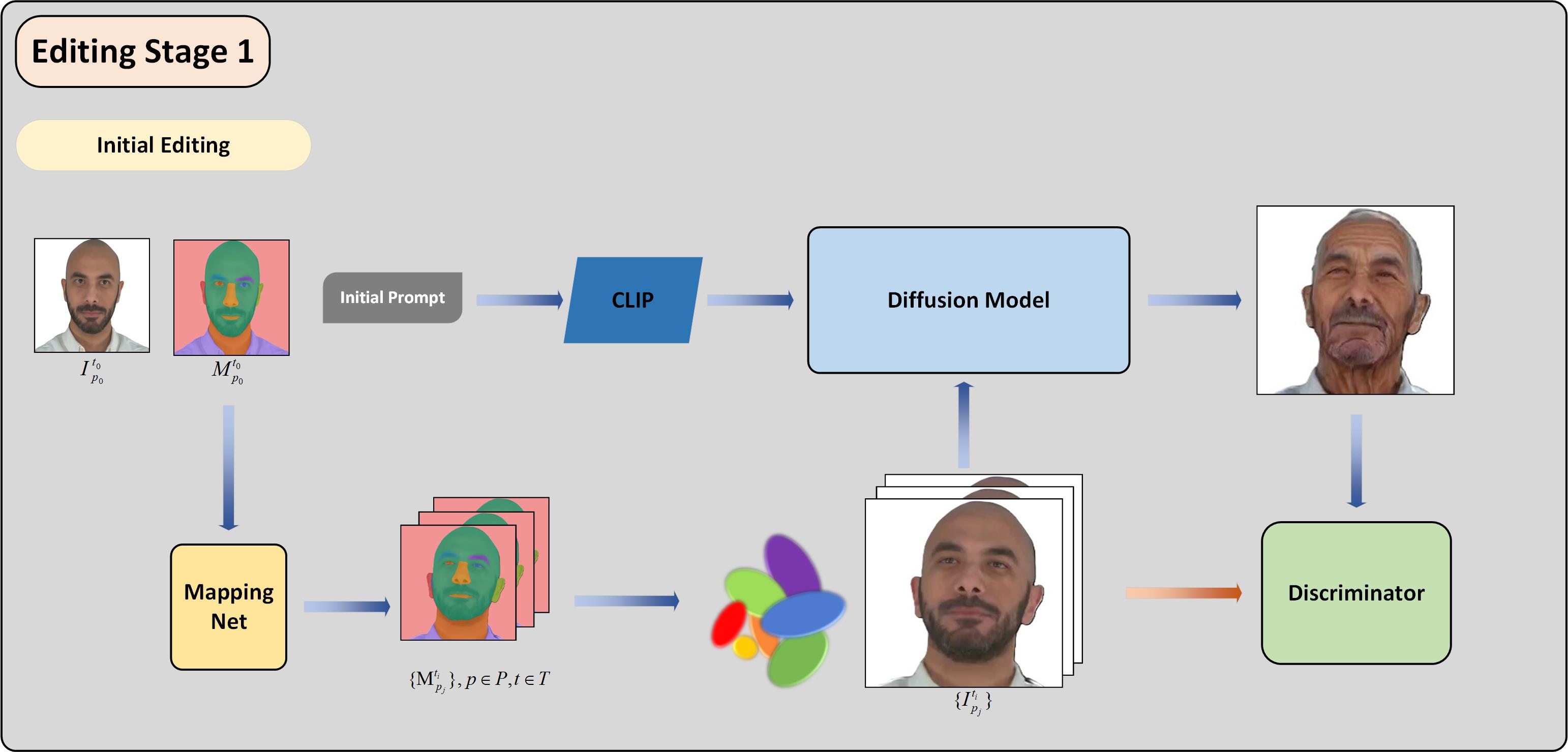}
    \caption{\textbf{Details of the First Editing Stage} For the image $I_{{{p}_{0}}}^{{{t}_{0}}}$ at the baseline time point ${{t}_{0}}$ and baseline camera pose ${{p}_{0}}$, the goal is to edit the selected masked area. To achieve this, we first warp the mask $M_{{{p}_{0}}}^{{{t}_{0}}}$ to $M_{{{p}_{j}}}^{{{t}_{i}}}$ using a mapping network. This step identifies Gaussian splats contributing to the masked area across different time points ${t_i}$ and camera poses ${p_j}$. Subsequently, we render and edit the corresponding images, ensuring the overall quality is preserved with the assistance of a discriminator. In the actual case, we use DALL-E as our conditional diffusion model.}
    \label{fig:stage_2_1}
\end{figure*}

To achieve flexible editing of expressions, textures, and additional accessories on head avatars, it is essential to use a technique that can both reconstruct an accurate head model and facilitate easy editing. Gaussian Avatars proposed using the FLAME\cite{FLAME:SiggraphAsia2017} mesh to model head avatars, enabling expressive changes. However, this method lacks control over accessories, such as hair, rings, or hats, which are not modeled by FLAME.

As shown in Figure.\ref{fig:stage_1}, we introduce a novel mesh-Gaussian binding method that differs from Gaussian Avatars \cite{qian2024gaussianavatars}. In our approach, we introduce two Gaussian tracking patterns for this stages of the process. The input video is processed with a photometric head tracker to fit the FLAME parameters. Each frame includes multi-view observations, timestep parameters, and known camera parameters. Initially, we track Gaussian splats across each triangle, similar to the method in Gaussian Avatars, to ensure that areas with significant changes can be modeled with high precision. Next, we apply an independent facial composition identifier to generate a semantic mask. This allows us to assign semantic labels to each Gaussian splat when rendered into images, ensuring that the same splats are tracked and manipulated consistently throughout the dynamic scene, maintaining temporal consistency during the editing process. Meanwhile, the rendered results are compared to the real images to train the avatar. In the next stage, we decouple the relationship between the Gaussian splats and the FLAME mesh, enabling the addition and modification of accessories, such as rings and hats.

To enhance rendering quality, we apply adaptive density control operations to adjust the density of the Gaussian splats, selectively densifying and pruning them as needed. We labels the new generated splats and track them to ensure that new Gaussian splats remain connected to the FLAME mesh and semantic masks, preserving the overall structure. Additionally, we optimize the position and scaling of the splats to improve the quality at the same time.

\subsection{Dynamic Gaussian Editing}

Traditional 3D editing \cite{lin2021deep} approaches rely on static 2D or 3D masks to restrict changes to specific regions. However, this approach presents challenges, as dynamic updates during training can render static masks inaccurate, limiting their effectiveness. Additionally, static masks in NeRF editing constrain changes to fixed spatial regions, which hampers natural content expansion beyond the mask boundaries. 

Recently, semantic segmentation-based masks were introduced by GaussianEditor \cite{chen2023gaussianeditor}, which partly address this issue by directly assigning semantic labels to individual Gaussian points. This technique increases the accuracy and speed of the editing process compared to previous approaches. However, applying these masks to dynamic scenes directly introduces new challenges in editing stage. For a labeled splat at time ${t_0}$, it may not contribute to the color of the corresponding semantic region at time ${t_1}$. Conversely, unlabeled splats could still influence the rendering process at later times. This inconsistency can lead to problems such as color mismatches, which modify the appearance of the head avatar, especially when the avatar undergoes significant pose changes over time.

\begin{figure}[htbp]
    \centering
    \includegraphics[width=0.3\textwidth]{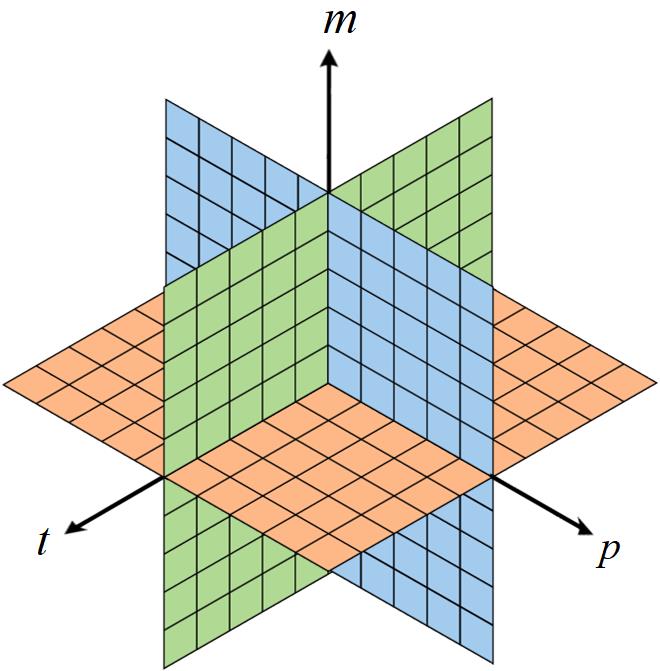}
    \caption{Space of our mapping net. It takes target timestep $t$, camera pose $p$ and original mask image $M_{{{p}_{0}}}^{{{t}_{0}}}$ as input, and output the target mask $M_{{{p}}}^{{t}}$. In order to keep the continuity of time and camera pose, we will apply bilinear interpolation on the plane of $tOp$. We train this module by utilizing the mask of the training dataset at different time and poses generated at the first stage. }
    \label{fig:triplane}
\end{figure}

To address this issue, we propose a method, illustrated in Figure.\ref{fig:stage_2_1}, that aims to consider all Gaussian splats that contribute to the results across different times and poses. Our approach identifies the shape of the desired editing mask at different times and camera poses. By using the mapping net that generate selected areas across the entire timeline, we can track the Gaussian splats contributing to the target area throughout the dynamic scene. Details of the mapping net can be find in Figure.\ref{fig:triplane}.

Next, we edit each image in the selected set to produce an edited image set. Finally, we apply a learning process with a conditional adversarial loss, which helps regulate the Gaussian splats and maintain temporal consistency. This method allows us to edit the entire dynamic model, incorporating the desired changes arbitrarily and efficiently.

\subsection{LLM-based Fine Editing}

\begin{figure*}[htbp]
    \centering
    \includegraphics[width=\textwidth]{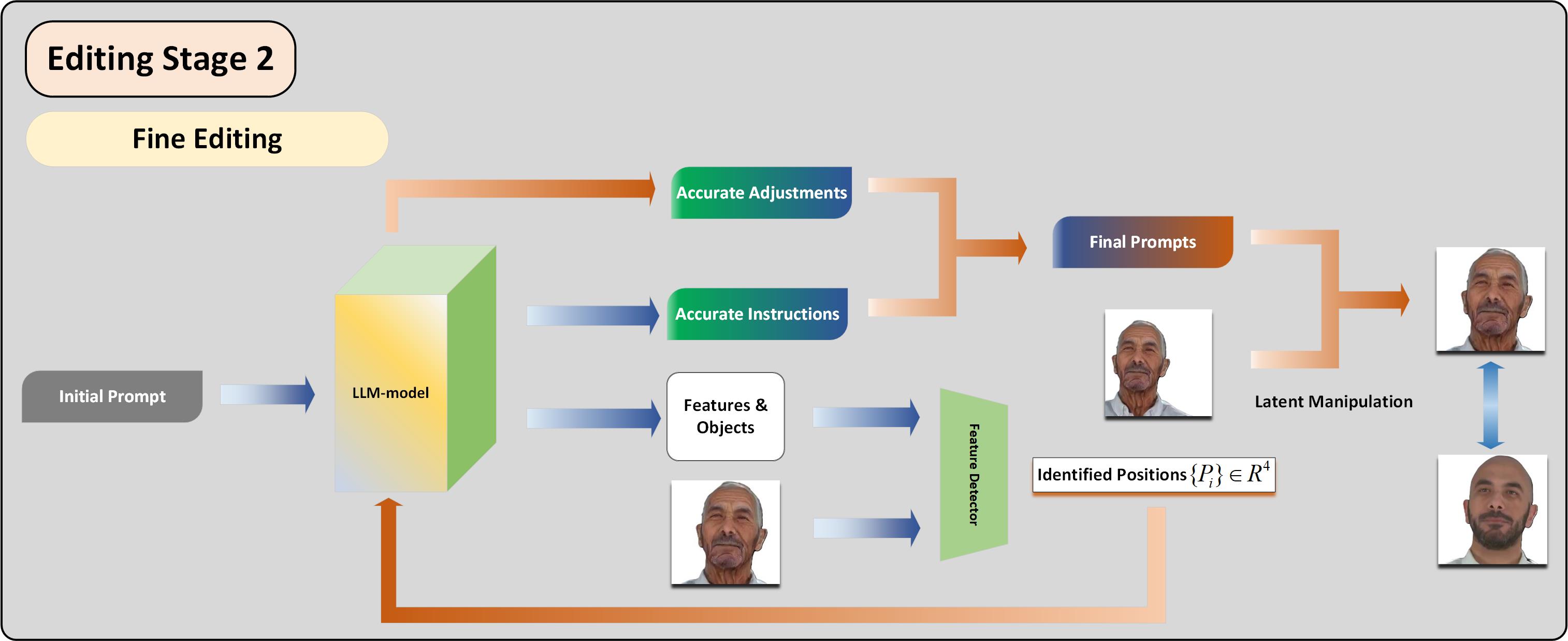}
    \caption{\textbf{Details of the Second Editing Stage} In this stage, the focus shifts to prompt manipulation and fine-grained editing. The process begins with taking the initial prompt as input. An LLM (Large Language Model) is utilized to refine the prompt into precise instructions while identifying the features and objects described within it. Next, a mapping net is employed to accurately locate the positions of the relevant elements in the images. The refined instructions, combined with the LLM's enhanced understanding, help generate the final, detailed prompts for editing. Finally, latent manipulation is applied to modify the representation, resulting in the ultimate edited images. In our experiment, we utilize GPT-4 as the LLM-model.}
    \label{fig:stage_2_2}
\end{figure*}

Previous 3D scene editing works based on diffusion model \cite{pandey2023diffusionhandlesenabling3d} have difficulties in retain the stability of edit and relatively low comprehension ability when facing extreme detailed prompts including described information like directions and relative positions. Works like SLD \cite{wu2023self} proposes a LLM-controlled self-correcting model that analyzes the components of prompts and execute instructions in order in latent space. However, this method still lacks the ability of understanding direction and relative position information. Work like \cite{bartal2023multidiffusionfusingdiffusionpaths}, bring up a feasible direction to utilize the LLM model to assistant the fine editing of images. 

In order to improve the generate quality of results when facing such difficult conditions, we focus on solving misplaced and misunderstand problems related to edit and add accessories based on precise detailed prompts.
As Figure.\ref{fig:stage_2_2} shows our method, we propose a framework similar to SLD which can provide a practical way of fine editing. We tend to readjust the structure of prompts based on the LLM and then carefully modify the images generated from the previous stage. This image correction is based on latent space manipulations, and it contains the alignment of multi-view consistency in our method.

\subsection{Loss Function and Regularization}

The primary loss should focus on the rendered images. Hence we deploy a color loss as follows:

\begin{equation}\label{eqn-11} 
{{L}_{rgb}}=\lambda {{L}_{2}}(I,\hat{I})+(1-\lambda ){{L}_{lpips}}(I,\hat{I})
\end{equation}

It should be noticed that this loss is used under different $\lambda $ value whenever we have the groundtruth image and need to learn the image-level information. In our experiment, we set $\lambda $ equals to 0.9 during our modeling stage and 0.7 during editing stage, it has been tested to be effective during ablation study in our work. 

The second constraint we need to pay attention is the tracking loss. This one concentrates on dealing with the relative position between the mesh and Gaussian splats, and affiliation between the certain semantic area and Gaussian splats. 

\begin{equation}\label{eqn-12} 
{{L}_{tracking}}=\lambda {{L}_{2}}(\overline{x},\Delta x)+(1-\lambda ){{L}_{perceptron}}(l,\widehat{l})
\end{equation}

To supervise the positions and distributions of the Gaussian splats during the editing stage in order to conserve the basic structure of the model, we need to employ a loss that can punish the misplacement of the splats. Meanwhile, we also need to be aware of optimizing the physical parameters of each Gaussian splat.

\begin{equation}\label{eqn-13} 
{{L}_{gs}}=\sum\limits_{i\in \{p,t,c\},j\in \mathbb{S}}{{{\lambda }_{i}}{{L}_{2}}(G{{({{x}_{j}})}_{i}},G{{({{\widehat{x}}_{j}})}_{i}})}
\end{equation}

where ${G({{x}_{j}})}$ represents Gaussian splat $j$, $i$ represents corresponding sub property of $j$ like positions, translations and color. For different parameters, we apply the coefficient ${{\lambda }_{i}}$ independently.
Finally, we define the adversarial loss in editing stage. The GAN method is used to reduce the color difference between the original one and the objective images.

\begin{equation}\label{eqn-14} 
{{L}_{D}}=E(max(0,1-D(x)))+E(max(0,1+D(G(z))))
\end{equation}

\begin{equation}\label{eqn-15} 
{{L}_{G}}=-E(D(G(z)))
\end{equation}

The overall loss function should be:

\begin{equation}\label{eqn-16} 
{{L}_{rec}}=\lambda {{L}_{rgb}}+(1-\lambda ){{L}_{tracking}}
\end{equation}

\begin{equation}\label{eqn-17} 
{{L}_{edit}}=\frac{{{\lambda }_{1}}{{L}_{rgb}}+{{\lambda }_{2}}{{L}_{gs}}+{{\lambda }_{3}}{{L}_{G}}}{{{\lambda }_{1}}+{{\lambda }_{2}}+{{\lambda }_{3}}}
\end{equation}

In our experiments, we use the Adam algorithm as the optimizer, setting the learning rate to ${1 \times 10^{-3}}$ at the beginning of the modeling stage, which gradually decreases to ${1 \times 10^{-5}}$. Each head avatar is trained independently from a scratch model, and we evaluate the fine-tuning ability and the effects of various editing methods on the distinct models constructed.

For each training session, the entire reconstruction process takes approximately 1 hour to complete. Densification and splitting frequencies are determined every 2048 iterations. However, we have not yet investigated the optimal timing for densifying and splitting the Gaussian splats. We believe that future work could involve developing a module capable of detecting Gaussian splat areas and automatically adjusting the densification and splitting processes to optimize performance.

\section{Experiments}

\subsection{Environment Settings}
Training:
We implement our benchmark using the NeRSemble dataset, where each subject has 16 different camera views around the human head. The model is trained with the same parameter settings and methods as GaussianAvatars, using two RTX 4090 GPUs.

Evaluation:
We evaluate our approach based on the following settings:

Reconstruction Evaluation: To demonstrate that our method maintains high reconstruction quality, we compare our results with those of other models.
Editing Evaluation: We assess our model's editing ability from various perspectives, including:
1) Driving the avatar with novel poses and expressions,
2) Modifying the style and identity of the head avatar based on prompts,
3) Adding accessories using prompts that contain detailed position and style information.

We conduct experiments and compare our results with baseline models for both self-reenactment and cross-identity reenactment tasks. For reconstruction and non-prompt-based editing metrics, we use PSNR, SSIM, and LPIPS to evaluate the visual quality of the rendered images. For prompt-based editing, we evaluate edit accuracy using SLM and text-image direction similarity with the InstructPix2Pix metric.

\begin{figure*}[htbp]
    \begin{subfigure}{\textwidth}
        \centering
        \includegraphics[width=\textwidth]{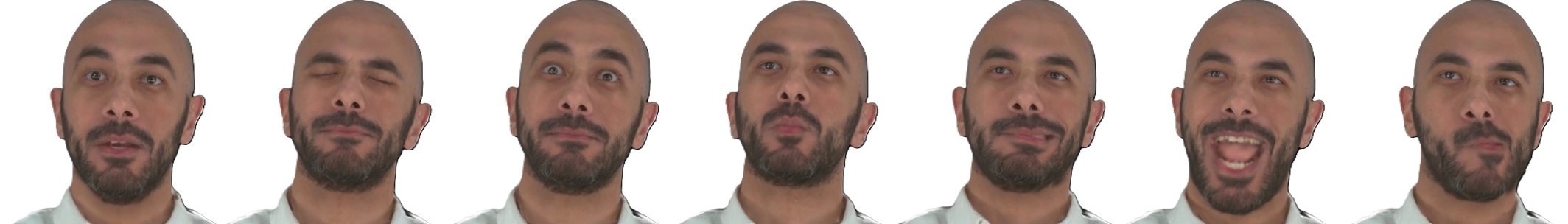}
        \caption{}
        \label{fig:demo_a}
    \end{subfigure}
    \quad
      \begin{subfigure}{\textwidth}
        \centering
        \includegraphics[width=\textwidth]{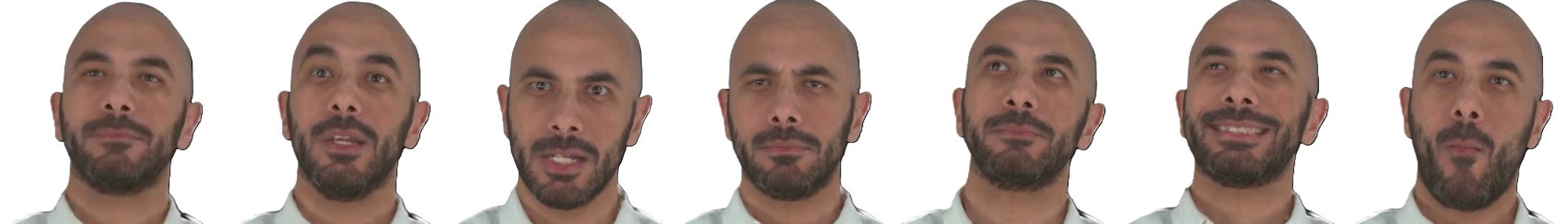}
        \caption{}
        \label{fig:demo_b}
    \end{subfigure}
    \caption{Demonstration of our expression editing experiments.}
    \label{experiments_1}
\end{figure*}

\begin{figure*}[htbp]
    \begin{subfigure}{\textwidth}
        \centering
        \includegraphics[width=\textwidth]{images/Original.jpg}
        \caption{Expression Editing with original appearance.}
        \label{fig:demo_0}
    \end{subfigure}
    \quad
    \begin{subfigure}{\textwidth}
        \centering
        \includegraphics[width=\textwidth]{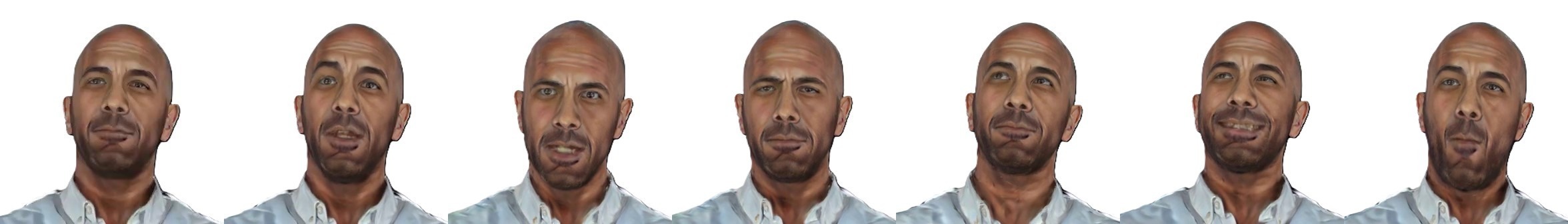}
        \caption{\textbf{Prompt}: Turn the man into Vin Diesel.}
        \label{fig:demo_1}
    \end{subfigure}
    \quad
    \begin{subfigure}{\textwidth}
        \centering
        \includegraphics[width=\textwidth]{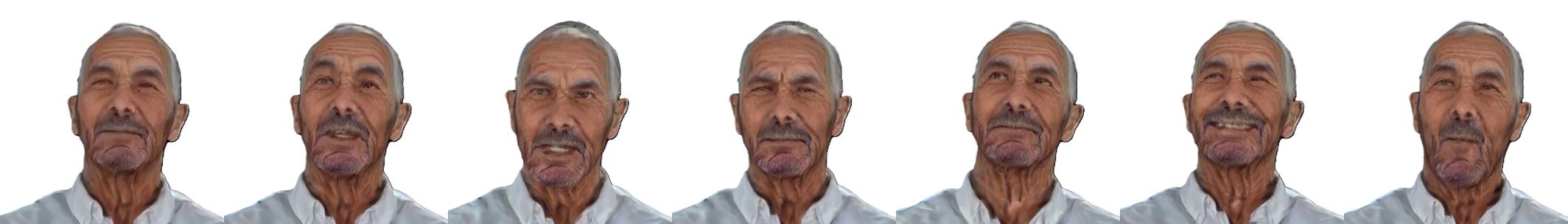}
        \caption{\textbf{Prompt}: Turn the man older.}
        \label{fig:demo_2}
    \end{subfigure}
    \quad
    \begin{subfigure}{\textwidth}
		\centering
		\includegraphics[width=\textwidth]{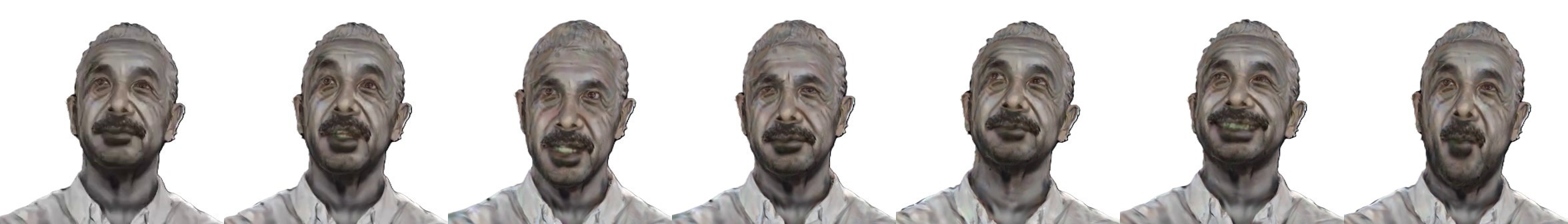}
		\caption{\textbf{Prompt}: Turn the man into Einstein.}
		\label{fig:demo_3}
    \end{subfigure}
    \quad
    \begin{subfigure}{\textwidth}
		\centering
		\includegraphics[width=\textwidth]{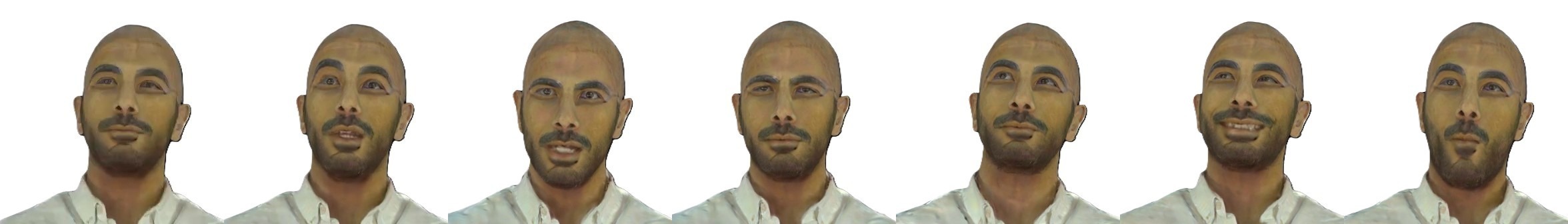}
		\caption{\textbf{Prompt}: Turn the man into Egyptian}
		\label{fig:demo_4}
    \end{subfigure}
    \caption{Demonstration of our style editing experiments.}
    \label{experiments_2}
\end{figure*}

\begin{figure*}[htbp]
    \centering
    \includegraphics[width=\textwidth]{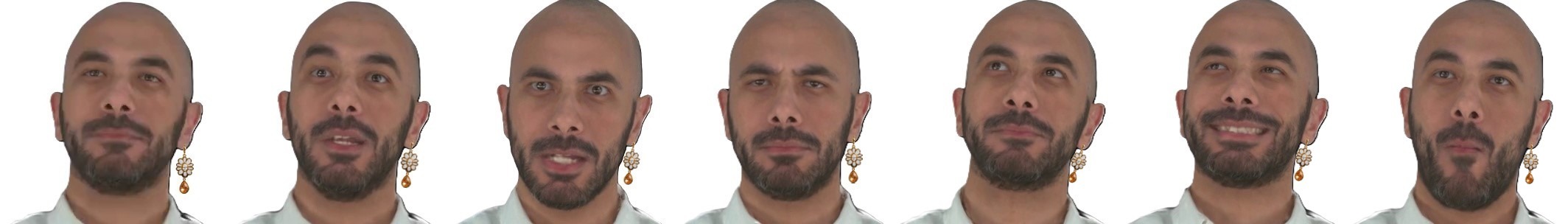}
    \caption{\textbf{Prompt}: Make the man's left ear wear an earring.\\Demonstration of our accessories editing experiments.}
    \label{experiments_3}
\end{figure*}

\subsection{Comparisons for Head Avatar Reconstruction}

\begin{table}[hbp]
  \centering
  \begin{tabular}{cccc}
    \toprule
    Metrics & PSNR$\uparrow$ & SSIM$\uparrow$ & LPIPS$\downarrow$ \\
    \midrule
    GaussianAvatars & 29.3 & 0.865 & 0.113  \\
    HeadStudio & 28.4 & 0.771 & 0.159  \\
    PointAvatar\cite{zheng2023pointavatardeformablepointbasedhead} & 27.1 & 0.689 & 0.147  \\
    Ours & \textbf{30.1} & \textbf{0.907} & \textbf{0.094}\\
    \bottomrule
  \end{tabular}
  \caption{Comparison of reconstruction quality. Ours is better.}
  \label{tab:metrics_1}
\end{table}

Figure.\ref{experiments_1} presents qualitative comparisons of novel-view synthesis from various methods. Our approach demonstrates exceptional performance compared to the listed works. Notably, GaussianEditor produces images with noticeable artifacts around the mouth and eye regions, and struggles to generate avatars with novel expressions.

This can be attributed to two key factors:
1) GaussianEditor does not utilize a mesh model for constraints, which limits the regularization of the Gaussian splat distribution.
2)The editing algorithm in GaussianEditor is prompt-based, and such edits lack the necessary information to effectively alter the expressions of head avatars. Additionally, the inherent randomness of the diffusion model makes it challenging to maintain stable control over fine-grained edits, such as expressions.

As shown in Table.\ref{tab:metrics_1}, we present quantitative comparisons between our method and several other prominent approaches.

\subsection{Comparisons for Head Avatar Editing}

\begin{table}[hbp]
  \centering
  \begin{tabular}{cccc}
    \toprule
    Metrics & Average & Spacial & Attribute \\
    \midrule
    GaussianEditor & 24.4\%  & 15.2\% & 33.6\% \\
    GenN2N & 36.4\% & 27.4\% & 45.4\% \\
    Ours & \textbf{67.5}\% & \textbf{64.9}\% & \textbf{70.1}\%\\
    \bottomrule
  \end{tabular}
  \caption{Comparison of edit quality. Ours is better.}
  \label{tab:metrics_2}
\end{table}

As illustrated in Figure.\ref{experiments_2} and Figure.\ref{experiments_3}, DynamicAvatars demonstrates outstanding scene editing and the ability to render finely edited images compared to other methods. It is evident that our model effectively interprets the information described in the prompt and generates photorealistic images, thanks to the preprocessing module. Moreover, our model excels in style modification while maintaining accurate expressions and texture colors, owing to the GAN framework that supervises facial image generation.

Table.\ref{tab:metrics_2} presents the statistics of the quantitative comparisons regarding editing ability. Our approach outperforms other methods in handling complex editing scenarios, such as detailed prompts and mixed style-expression edits.

\section{Conclusion}

We propose DynamicAvatars, which enhances control and flexibility in editing. The dual tracking of Gaussians enables improved reconstruction and editing quality, while the prompt preprocessing architecture enhances the diffusion model’s ability to generate accurate edited images. Additionally, the incorporation of a GAN method helps reduce color discrepancies, making style edits more natural, particularly in facial regions. Furthermore, our approach integrates various functionalities for manipulating head avatars and optimizes each step to achieve higher quality results. The dynamic Gaussian editing capability allows for more efficient and intuitive editing of dynamic scenes.

{
    \small
    \bibliographystyle{ieeenat_fullname}
    \bibliography{main}
}

\end{document}